\documentclass[fleqn,usenatbib]{mnras}

\usepackage[T1]{fontenc}
\usepackage{ae,aecompl}

\usepackage{graphicx}	
\usepackage{amsmath}	
\usepackage{amssymb}	
\usepackage{color}

\newcommand{\rein}{$R_{\rm Ein}$}
\newcommand{\rhalf}{$r_{1/2}$}
\newcommand{\mathbbm}[1]{\text{\usefont{U}{bbm}{m}{n}#1}}

\title[Machine learning and quasar microlensing]{Quasar microlensing light curve analysis using deep machine learning}

\author[G. Vernardos \& G. Tsagkatakis]{
Georgios Vernardos,$^{1}$\thanks{E-mail: gvernard@astro.rug.nl (GV)}
and Grigorios Tsagkatakis$^{2}$
\\
$^{1}$Kapteyn Astronomical Institute, University of Groningen, PO Box 800, NL-9700AV Groningen, the Netherlands\\
$^{2}$Institute of Computer Science - Foundation for Research and Technology (FORTH), GR-71110, Heraklion, Greece
}

\date{Accepted XXX. Received YYY; in original form ZZZ}

\pubyear{2019}

\begin{document}
\label{firstpage}
\pagerange{\pageref{firstpage}--\pageref{lastpage}}
\maketitle

\begin{abstract}
We introduce a deep machine learning approach to studying quasar microlensing light curves for the first time by analyzing
hundreds of thousands of simulated light curves with respect to the accretion disc size and temperature profile.
Our results indicate that it is possible to successfully classify very large numbers of diverse light curve data and measure the accretion disc structure.
The detailed shape of the accretion disc brightness profile is found to play a negligible role, in agreement with \cite{Mortonson2005}.
The speed and efficiency of our deep machine learning approach is ideal for quantifying physical properties in a `big-data' problem setup.
This proposed approach looks promising for analyzing decade-long light curves for thousands of microlensed quasars, expected to be provided by the Large Synoptic Survey Telescope.
\end{abstract}


\section{Introduction}
Quasar microlensing is a unique way of measuring the structure of AGN at the very small scales inaccessible by other techniques.
The magnification effect of the microlenses - stellar mass compact objects lying within lensing galaxies - on small angular diameter sources, like an accretion disc surrounding a supermassive black hole, can be very pronounced, allowing one to study their structure on $10^{-6}$ arcsec scales, corresponding to a few tens of AU on the source plane \citep[see the review by][]{Schmidt2010}.
Such measurements are putting classic accretion disc models, particularly the standard thin disc model \citep{Shakura1973}, to the test \citep{Abolmasov2012b}: quasar accretion discs are found to be larger than expected \citep{Morgan2010,Blackburne2011,Jimenez2014,Chartas2016} while their temperature profiles range from shallower \citep{Bate2008,Rojas2014,MacLeod2015,Bate2018} to steeper \citep{Eigenbrod2008,Floyd2009,Blackburne2011,Jimenez2014}.

Microlensing snapshot and monitoring observations can produce single-epoch or time series of observations (light curves).
Light curve analysis techniques allow one to extract physical information from year-long observations, at the cost of higher model complexity and much more demanding computations.
The method of \cite{Kochanek2004} provides a powerful Bayesian analysis approach to studying quasar structure by generating $\mathcal{O}(10^6)$ simulated light curves for each combination of model parameters and fitting them to the data \citep{Poindexter2008}.
This computationally demanding technique requires careful use of priors (e.g. on the source effective velocity components), and may have to resort to independent modelling of light curve segments to keep the computations feasible \citep{Poindexter2010b}.

Microlensing analyses are based on some knowledge of the mass-density distribution of the lensing galaxy and its environment (the macromodel), which depends on the quality of the imaging data, the presence of satellite galaxies, possible existence of dark matter substructures, etc.
This model determines the values of the convergence, $\kappa$, and shear, $\gamma$, at the locations of the multiple quasar images, which in turn set the microlensing properties via the magnification map \citep{Paczynski1986,Kayser1986}.
If these parameters, or the further partition of the convergence into smooth (dark) and compact (stellar) matter components \citep[e.g.][]{Schechter2002}, are not accurately known, the resulting microlensing effect can vary significantly \citep{Vernardos2014c}.
Factoring in such uncertainties in current microlensing light curve analysis techniques quickly leads to intractable computations.

The number of known quasar lenses is set to increase from a few hundreds to thousands by all-sky survey facilities currently under construction.
The Large Synoptic Survey Telescope \citep[][]{LSST2009} will provide multi-wavelength light curves for all these lenses over a period of a decade or more.
Analyzing this large volume of data will require new, more efficient methods than the current state-of-the-art techniques, which have been designed to perform optimally for single systems with high quality observations.

Machine learning algorithms have provided promising solutions to `big-data' problems in astronomy \citep[e.g.][]{Hala2014,HuertasCompany2015,Tuccillo2018,Stivaktakis2018}.
Convolutional Neural Networks \citep[CNNs][]{LeCun2015} are a particular class of such algorithms that can model complicated non-linear relationships by repeated abstract processing layers.
CNNs have been successfully used in gravitational lensing to find lenses \citep{Lanusse2018,Petrillo2019} and model them \citep{Hezaveh2017}.

In this paper, we use CNNs to model quasar microlensing light curves and extract the size and temperature profile of the accretion disc.
Our training set, based on simulated light curves, and the neural network architecture are presented in Section 2.
Section 3 consists of the training results and a series of more advanced tests on different types of light curves.
Conclusions and implications for future work are discussed in Section 4.

Throughout this paper a cosmological model with $\Omega_{\rm m}=0.26$, $\Omega_{\rm \Lambda}=0.74$, and $H_0=72$ km s$^{-1}$ Mpc$^{-1}$ is used.

\section{Method}
\label{sec:method}
There are two main factors affecting the performance of CNN algorithms: the training set and the design of the network itself.
Here we present the steps in generating mock quasar microlensing light curves, used to create the training set and a series of additional test sets, and the main features of the adopted CNN architecture.

\subsection{Generating mock light curves}
Generating light curves follows the same approach used in previous work \citep{Vernardos2015} and elsewhere in the literature \citep[e.g.][]{Kochanek2004}.
The necessary components for this procedure are: the accretion disc brightness profile, the Einstein radius of the microlenses, the magnification map, and a velocity model for the accretion disc moving across it.
Our assumptions on the size and shape of the accretion disc profile, which are the main focus of this work, and the rest of the parameters involved are described below.

The size of the accretion disc as a function of wavelength is assumed to have the following parametric form:
\begin{equation}
\label{eq:parametric}
r = r_0 \left( \frac{\lambda}{\lambda_0}\right)^{\nu},
\end{equation}
where $\lambda = \lambda_{\rm obs}/(1+z_S)$, with $\lambda_{\rm obs}$ being the observing wavelength and $z_S$ the quasar redshift, $\nu$ the power law index, and $r_0$ the size of the disc observed at the rest wavelength $\lambda_0 = 102.68$ nm.
This is a frequently used parametric model \citep[e.g.][]{Jimenez2014,Bate2018} that captures the accretion disc size dependence on physical parameters like the Eddington luminosity, the accretion efficiency, and the black hole mass, in $r_0$.
Temperature is inversely proportional to wavelength for accretion disc emission that is a superposition of black body spectra.
This is the case for the thin disc model that has $\nu = 4/3$.

\begin{figure}
	\includegraphics[width=0.5\textwidth]{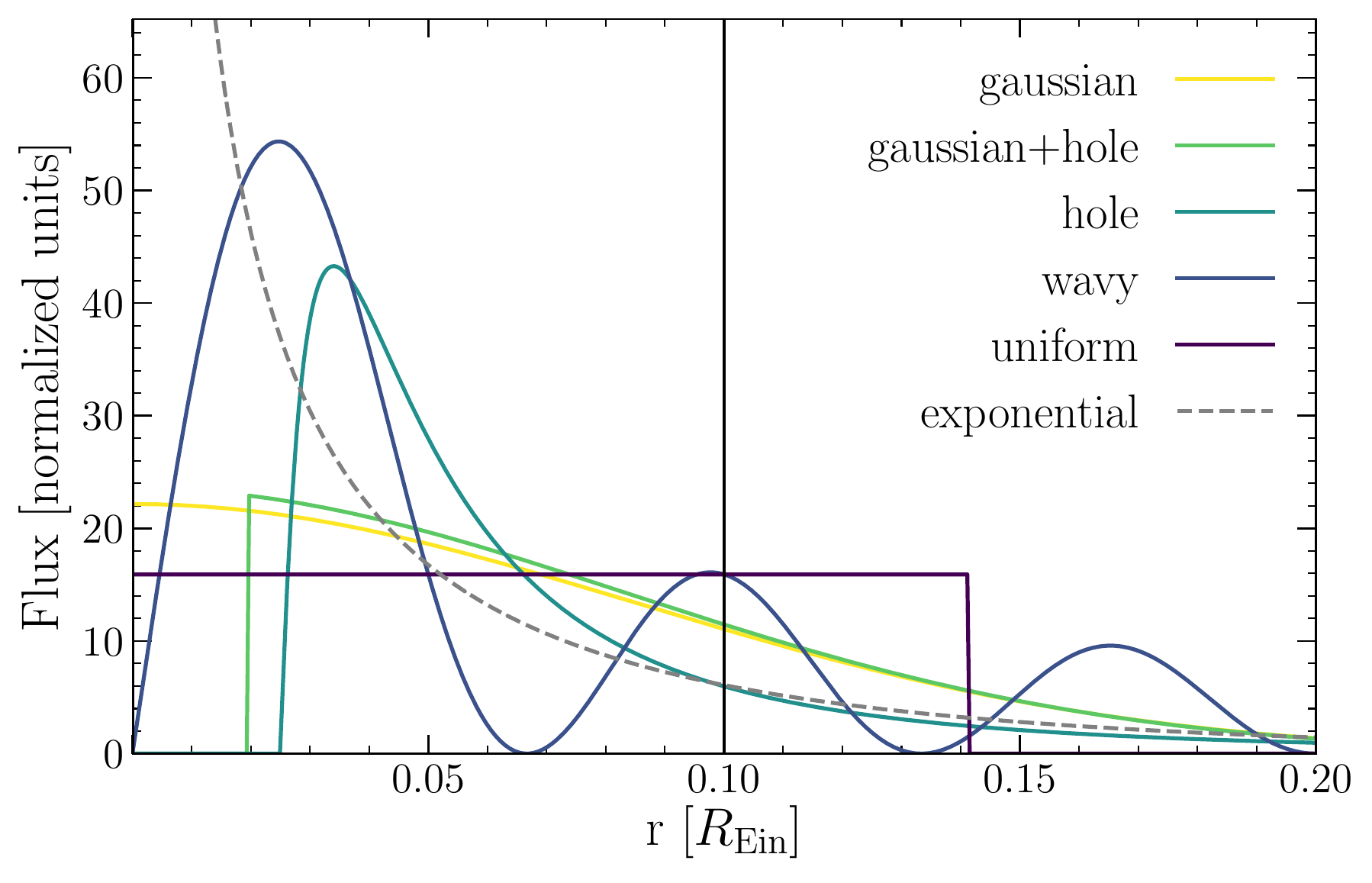}
	\caption{One-dimensional intersections of the accretion disc surface brightness profiles listed in Table \ref{tab:shapes}, in units of the Einstein radius (equation \ref{eq:rein}) that is set to $4.64 \times 10^{16}$ cm. All the profiles have been scaled to have the same half-light radius, \rhalf$ = 0.1$ \rein~(vertical line). The inner edge of the `gaussian$+$hole' profile corresponds to the innermost stable circular orbit of a $10^9$ M$_{\odot}$ black hole ($9\times 10^{14}$ cm), and the `wavy' profile has $n=3$. The `exponential' profile was not used in the training of our classification algorithm, but it was used for tests (see Section \ref{sec:expo}).}
	\label{fig:profiles}
\end{figure}

The half-light radius, \rhalf, which contains half the luminosity of the accretion disc, is set to the size obtained from equation (\ref{eq:parametric}).
Although there is evidence that the shape of the disc profile does not play a significant role in microlensing observations \citep{Mortonson2005}, this has not been examined thoroughly with respect to light curves, and not at all as a function of $\kappa,\gamma$.
Here, we assume different fiducial parametric shapes aiming to cover as wide a range of brightness variations as possible.
The different brightness profiles have the following shapes: gaussian, gaussian with an inner edge (`gaussian$+$hole'), smooth with an inner edge (`hole'), sinusoidal dropping with radius (`wavy'), uniform, and exponential.
These are all symmetric two-dimensional profiles so that the brightness per unit surface, $I(r)$, is only a function of the radius (see Table \ref{tab:shapes}).
Their one-dimensional intersections are shown in Fig. \ref{fig:profiles}.
Apart from the size, each profile may have a different number of free parameters, which are listed in Table \ref{tab:shapes}.
All the profiles are truncated at $2 \times$ \rhalf~(the full profiles will only induce a small magnification offset due to their flatness beyond $2 \times$ \rhalf) and normalized to a luminosity of unity.
\begin{table*}
	\centering
	\caption{Shapes of the accretion disc surface brightness profile, $I(r)$. The normalization factor corresponds to the luminosity, calculated as the two-dimensional integral of the surface brightness. Every shape has a parameter associated to the profile size, which depends on the half-light radius, \rhalf, the radius containing half the luminosity. The `gaussian$+$hole' profile has an inner edge, determined by $R_{\rm in}$ that is equal to the radius of the innermost stable circular orbit of a black hole with mass $M$ ($G$ is the gravitational constant and $c$ the speed of light). The `wavy' profile has an additional parameter $n$, which is the number of nodes of the profile within two half-light radii. The `exponential' profile was not used in the training set (Section \ref{sec:training}) but it was used to test the performance of our classification method on data that it was not trained on (Section \ref{sec:expo}).}
	\label{tab:shapes}
	\begin{tabular}{rcccc}
		name 			& analytic form	of $I(r)$	& normalization factor	& size parameter 			& other parameters 	\\
						&							&						& (function of \rhalf)		&					\\
		\hline
			\rule{0pt}{15pt}		gaussian 		& $\exp[-\frac{r^2}{2\sigma^2}]$ 								& $\frac{1}{2\pi\sigma^2}$ 		& $\sigma = \frac{r_{1/2}}{1.18}$ 	& - \\
			\rule{0pt}{15pt}		gaussian$+$hole	& $\begin{cases} 0 & \text{for } r \leq R_{\rm in} \\ \exp[-\frac{r^2}{2\sigma^2}] & \text{for } r > R_{\rm in} \end{cases}$ & $\frac{1}{2\pi \sigma^2} \, \exp[\frac{R_{\rm in}}{2\sigma^2}]$ & $\sigma = \sqrt{\frac{r_{1/2}^2 - R_{\rm in}^2}{2 \, \mathrm{ln}2}}$ & $R_{\rm in} = \frac{6GM}{c^2}$ \\
			\rule{0pt}{15pt}		hole	 		& $\frac{1}{r^3} (1 - \sqrt{\frac{R_{\rm in}}{r}} )$ 			& $\frac{3 R_{\rm in}}{2\pi}$ 	& $R_{\rm in} = \frac{r_{1/2}}{4}$ 	& - \\
			\rule{0pt}{15pt}		wavy			& $\frac{\sin^2 (\beta r)}{r}$ 									& $\frac{\beta}{n \pi^2}$ 		& $\beta = \frac{n \pi}{2 \, r_{1/2}}$ & $n$ \\
			\rule{0pt}{15pt}		uniform			& 1																& $\frac{1}{\pi R^2}$ 			& $R = \sqrt{2} \, r_{1/2}$			& - \\
		\hline
			\rule{0pt}{15pt}		exponential		& $\left( \exp[(\frac{r}{\sigma})^{3/4}] -1 \right)^{-1}$		& $\frac{0.062}{\sigma^2}$		& $\sigma = 0.4106 \, r_{1/2}$		& - \\
		\hline
	\end{tabular}
\end{table*}

The scale length of microlensing is set by the Einstein radius of the microlenses on the source plane:
\begin{equation}
\label{eq:rein}
R_{\rm Ein} = \sqrt{\frac{4G\langle M \rangle}{c^2} \frac{D_{\mathrm{S}} \, D_{\mathrm{LS}}}{\mathrm{D_L}}},
\end{equation}
where $D_{\mathrm{S}},D_{\mathrm{L}},$ and $D_{\mathrm{LS}}$ are the angular diameter distances to the source, the lens, and between the lens and the source respectively, $\langle M\rangle$ is the average microlens mass, $G$ the gravitational constant, and $c$ the speed of light.
Here we use microlenses with 1 M$_{\odot}$ and the fiducial lens and source redshifts of 0.5 and 1.2, to get \rein $= 4.65 \times 10^{16}$ cm.
However, as it turns out only the relative values of \rhalf~with respect to \rein~are important and the absolute value of \rein~is irrelevant for the results presented here.

\begin{figure*}
	\includegraphics[width=\textwidth]{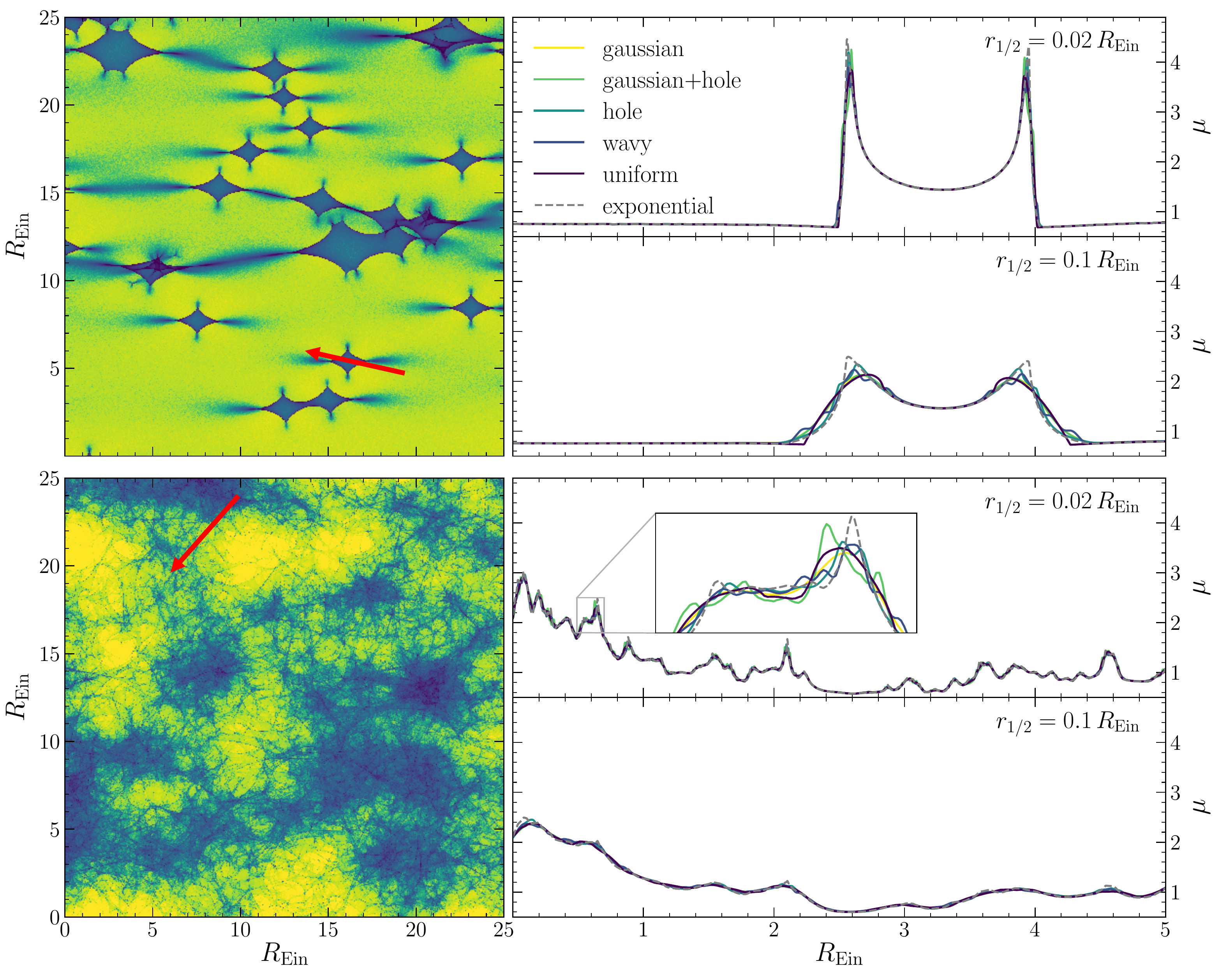}
	\caption{Mock light curves from magnification maps with $\kappa=0.09,\gamma=0.55$ (top) and $\kappa=0.8,\gamma=0.1$ (bottom). The light curve trajectories on the maps (red arrows) have the same length ($5\times$ \rein) and a random orientation. Mock light curves for all the profile shapes listed in Table \ref{tab:shapes} are shown for two different accretion disc profile sizes (\rhalf~equals to $=0.02$ and $=0.1$ \rein~respectively, for the top and bottom panels next to each map). The magnification, $\mu$, is with respect to the macromodel induced magnification, $\mu_{\rm macro} = [(1-\kappa)^2 - \gamma^2]^{-1}$, which is different for each map. Although the light curves may appear indistinguishable on large scales ($\sim$\rein), the effect of small scale profile structure is visible during a caustic-crossing event, as indicated in the inset plot in the third panel on the right-hand side.}
	\label{fig:lcurves_maps}
\end{figure*}

The effects of microlensing are simulated by magnification maps: pixellated representations of the source plane caustics due to the microlenses.
The main parameters\footnote{The smooth matter fraction, $s$, which is the fraction of smoothly distributed matter (as opposed to compact microlenses) over the total matter, is another important parameter for microlensing studies \citep[e.g.][]{Vernardos2014a}. Here we always set $s=0$ for simplicity. A varying $s$ has a similar effect to the magnification maps as changing the $\kappa,\gamma$, which is out of the scope of this work.} of these maps are the convergence, $\kappa$, and shear, $\gamma$, which determine the density and clustering of the caustic network (see Fig. \ref{fig:lcurves_maps}).
We use high quality magnification maps from the GERLUMPH\footnote{\url{http://gerlumph.swin.edu.au/}} parameter survey \citep{Vernardos2014a} that cover a large range of $\kappa,\gamma$ parameters.
These square maps are 25 \rein~wide, enough to allow for many independent and long light curves to be extracted, and have a 10,000 pixel resolution on a side, enough to resolve the details of the different disc profile shapes (e.g. the inner edge of the `gaussian$+$hole' profile, at $9\times10^{14}$ cm, corresponds to 8 pixels for \rein $ = 4.65 \times 10^{16}$ cm).

In order to extract microlensing observables for accretion discs that are not point-sources (\rhalf $>1$ pixel, or $> 0.0025$ \rein, which is always the case here), the magnification map and the disc profile need to be convolved first.
This leads to a reduced available map area, the effective map, due to the convolution edge effects.
The largest profiles used here are $\sim0.8$ \rein~wide, leaving an effective map of 24.2 \rein~(9,680 pixels) to extract light curves from.

The combination of the velocity vectors of the observer, the lens, and the source, results in an effective velocity of the source across the magnification map \citep[see][for a description of such a velocity model]{Kochanek2004}.
The length of a light curve on the magnification map is set by the magnitude of this velocity multiplied by the time range spanned by the observations (usually up to years).
The orientation of the map with respect to the effective velocity depends on the direction of the external shear, which can be derived from macromodels of the lensing galaxy.
However, these model components are more relevant to individual object studies and are outside the scope of this work that focuses on the size of the accretion disc with respect to \rein.
Therefore, the following broad assumptions are made: the light curve trajectories have a random direction, all light curves have the same length, and the light curve data are continuous.
An example of mock light curves is shown in Fig. \ref{fig:lcurves_maps}.

\subsection{Analysis}
The dramatic increase in the volume of astronomical observations has led to a situation where the manual analysis of highly dynamic observations by human experts is creating a bottleneck in the observation-to-knowledge pipeline.
To address this challenge, cutting-edge automated data analysis systems based on machine learning are gaining significant traction.
To that end, we consider deep machine learning approaches based on one-dimensional CNNs for supervised modeling of light curve data, where annotated examples are utilized for training multi-layer deep neural networks that can then characterize new observations.
We discuss the methodology for generating the training data set first and the architecture of the proposed CNN next. 

\subsubsection{Training set}
\label{sec:training}
Using the procedure described above, a set of mock light curves is generated for training the deep machine learning analysis algorithm.
The Einstein radius is set to $4.65 \times 10^{16}$ cm, corresponding to a lens and a source at redshifts 0.5 and 1.2 respectively, and 1 M$_{\odot}$ microlenses.
The half-light radius that sets the size of all the different accretion disc profiles varies in the range $0.02 - 0.2$ \rein.
The additional parameters for the `gaussian$+$hole' and `wavy' profiles are fixed to $R_{\rm in} = 9 \times 10^{14}$ cm and $n=3$ respectively (see Table \ref{tab:shapes} and Fig. \ref{fig:profiles}).
The different sizes are listed in Table \ref{tab:classes}.
The exponential profile is excluded from the training set and is used only for testing (Section \ref{sec:expo}).

In order to match the sizes to observed wavelengths, equation (\ref{eq:parametric}) is used with $\nu = 4/3$, the theoretical result from the thin disc model, and $r_0 = 2.59 \times 10^{15}$ cm, a somewhat smaller value than what is found from observations \citep[e.g. $r_0 = 1.38^{+0.45}_{-0.37} \times 10^{16}$ cm from][after converting their $r_s$ to \rhalf]{Jimenez2014}.
As discussed in Sections \ref{sec:results} and \ref{sec:conclusions}, however, the absolute scale of the problem is unimportant and only the \rhalf/\rein~ratio matters.

The training set is extracted from a single magnification map having $\kappa=0.09,\gamma=0.55$ ($s=0$), shown in the top left of Fig. \ref{fig:lcurves_maps}.
The length of the light curves is arbitrarily set to 5 \rein~($\approx$2,000 pixels\footnote{In fact, to produce a light curve we sample the magnification map along a given direction 2,000 times at intervals equal to the size of a pixel.}), which is long enough to permit several caustic crossings to occur, yet short enough to allow for a large number of independent light curves to be extracted from the 24.2 \rein-wide effective maps.

Thousands of light curves are generated for each profile and filtered to keep the ones having a maximum magnification of at least 1.5.
This is to avoid flat light curves without any meaningful features, which have a high chance to occur due to the large regions of the magnification map that are not covered by caustics (see top left panel of Fig. \ref{fig:lcurves_maps}).
For each \rhalf~and profile shape, $900$ light curves are selected for the training and $100$ for the validation sets, leading to a total of $45,000$ and $5,000$ respectively.

\begin{table}
	\centering
	\caption{Different sizes for the accretion disc profiles, parametrized by the half-light radius, \rhalf~(see also Table \ref{tab:shapes}). Each \rhalf~value is assigned to a class that is used to train the machine learning algorithm described in Section \ref{sec:nn}. A value of \rein$=4.64\times10^{16}$ cm is used and the corresponding observed wavelengths are calculated from equation (\ref{eq:parametric}) with $r_0=2.59\times10^{15}$ cm and $\nu=4/3$, for a source at redshift 1.2. However, as discussed in Sections \ref{sec:results} and \ref{sec:conclusions}, the absolute values of \rein, \rhalf, and the corresponding $\lambda_{\rm obs}$ do not affect the results of the analysis and are only listed for illustrative purposes.}
	\label{tab:classes}
	\begin{tabular}{rccc}
		class		& \rhalf~[\rein]	&	\rhalf~[$\times10^{14}$ cm]	&	$\lambda_{\rm obs}$ [nm]	\\
		\hline
		0 &    0.02 &    9.3 &  104.78 \\
		1 &    0.04 &   18.6 &  176.22 \\
		2 &    0.06 &   27.9 &  238.86 \\
		3 &    0.08 &   37.2 &  296.38 \\
		4 &    0.10 &   46.5 &  350.37 \\
		5 &    0.12 &   55.8 &  401.71 \\
		6 &    0.14 &   65.1 &  450.95 \\
		7 &    0.16 &   74.4 &  498.45 \\
		8 &    0.18 &   83.7 &  544.49 \\
		9 &    0.20 &   93.0 &  589.26 \\
		\hline
	\end{tabular}
\end{table}

\subsubsection{Convolutional Neural Network architecture}
\label{sec:nn}
Each one-dimensional light curve example is analyzed by a CNN architecture primarily consisting of a sequence of five layers-blocks, each one comprised of convolutional, non-linear activation, and max pooling layers, while the final predictions are produced by a sequence of fully connected layers.
The convolutional layer is responsible of convolving the input signals with a set of trainable kernels, $32$ in our case, of different temporal extent so that features of different temporal scale are identified.
More specifically, the one-dimensional light curve signal is first convolved by a set of $32$ unit-length kernels, which are then introduced as input to first layers-block.
The input to each layers-block is sequentially convolvoved with kernels of $10$, $20$ and $50$ spatial extend and passed through a non-linear Rectified Linear Unit (ReLU) activation function.
The signal produced by the last ReLU layer is added back to the original input following a residual modeling architecture \citep{He2016} and down-sampled by a max-pooling layer with pooling size equal to $3$ and stride of $2$.
The configuration and associated number of tunable parameters are shown in Table \ref{tab:cnn_arch}.  

The output of the five layers-blocks is subsequently passed through three fully connected layers of $128$, $32$ and $10$ units each, where the number of units in the final layer is dictated by the label resolution in our model (\rhalf~sizes in Table \ref{tab:classes}).
To make the actual prediction, the output of the final fully connected layer is introduced into a softmax layer which is responsible for producing probabilistic outputs (non-negative that sum to one), so that the class with the maximum probability corresponds to the predicted class.
Finally, in order to increase the generalization capacity of the proposed architecture, we employ two additional layers, a batch normalization layer at the end of each layers-block and dropout layers ($75\%$ dropout rate) between the fully connected layers. 

\begin{table}
\centering
\caption{Architecture of a layers-block component of our CNN.}
\begin{tabular}{|l|l|l|}
Type                & Shape   & Parameters \\ \hline \hline
Convolution         & (32,10) & 10272      \\ \hline
ReLU                & -       & -          \\ \hline
Convolution         & (32,20) & 20512      \\ \hline
ReLU                & -       & -          \\ \hline
Convolution         & (32,50) & 51232      \\ \hline
ReLU                & -       & -          \\ \hline
Add                 & -       & 0          \\ \hline
Pooling             & (3,2)   & 0          \\ \hline
Batch Normalization & -       & 0          \\ \hline
\end{tabular}
\label{tab:cnn_arch}
\end{table}

\begin{figure}
	\includegraphics[trim={0pt 10pt 0pt 0pt},clip,width=0.48\textwidth]{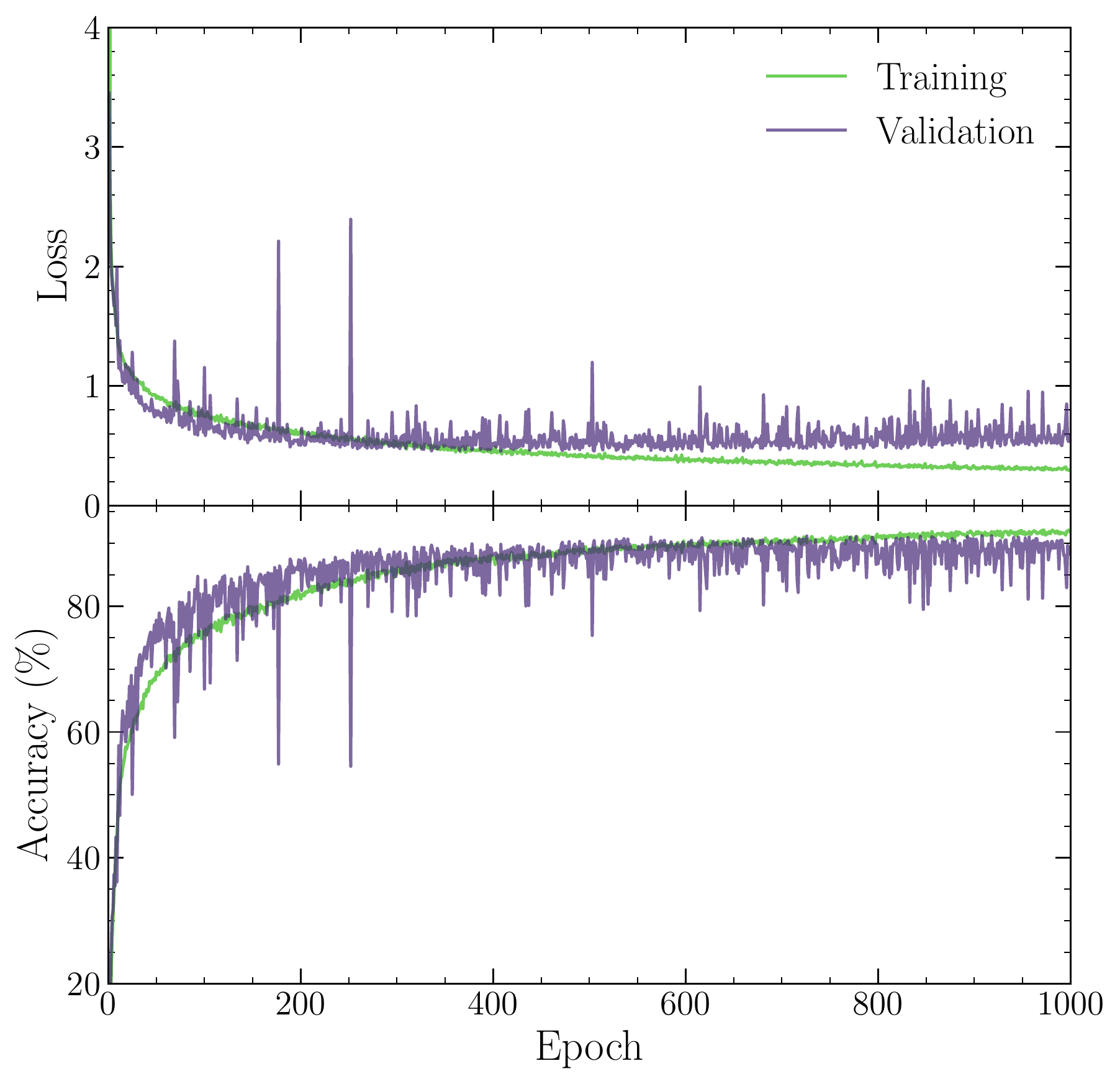}
	\caption{Value of the loss function (top) from equation (\ref{eq:cross_entropy}) and prediction accuracy (bottom, percentage of the correctly predicted \rhalf~classes over the total) as a function of epoch for the full training and validation sets. The network architecture used is the one highlighted in Table \ref{tab:acc_model}.}
	\label{fig:acc_loss}
\end{figure}

In total, the proposed CNN is parameterized by $673,098$ parameters that are tuned during the training stage to minimize the categorical cross entropy between predicted ($\hat{y}$) and actual labels ($y$).
For a single example classified in one out of $C$ classes, this is given by:
\begin{equation}
\label{eq:cross_entropy}
\mathcal{L}(y,\hat{y})=- \sum_{c=1}^{C}  \mathbbm{1}_{y_i \in C_c} \log p(\hat{y}_i),
\end{equation} 
where $\mathbbm{1}_{y_i \in C_c}$ is the indicator function taking a value of one only for the correct class $c$, and $p(\hat{y}_i)$ is the predicted probability associated with each class. 
Minimization of the loss function is achieved by employing the Adam optimizer \citep[][with learning rate $10^{-3}$, decay $10^{-5}$]{Kingma2014}.
To quantify the performance of the CNN architecture, we report the prediction accuracy as the percentage of correctly predicted \rhalf~classes, averaged over all examples (either training or validation, see Fig. \ref{fig:acc_loss}).
To maximize the accuracy, the training stage involves the repetitive tuning of parameters for $1000$ epochs, where each epoch corresponds to a single pass over the entire training set ($45,000$ light curves).

\section{Results}
\label{sec:results}
Here we present the results of the network training and its application.
First, we explore how modifying the neural network architecture can achieve better results with the training set.
Predictions can be only as good as the data used to train the algorithm, while a well-trained network is not guaranteed to give good predictions for data of a different kind.
This effect of the training/application data on the predictions is explored in a number of tests.
Finally, the ability to correctly recover a fiducial size-wavelength dependence of the accretion disc profiles is measured.
We point out that the network output is always one of the classes of \rhalf~listed in Table \ref{tab:classes}.

\subsection{Convergence}

To demonstrate the need for multi-layered CNN architectures, Table \ref{tab:acc_model} reports the achieved accuracy for a limited set of training and validation examples ($4500$ and $500$ respectively) at different epochs and for different architectures.
Specifically, we consider networks with $2$, $3$, $4$ and $5$ layers-blocks, while for the case of $5$ layers-blocks, we also explore the impact of drop-out (DO) regularization.
For each case, we report the accuracy achieved on the validation set (and the training set in parenthesis). 

\begin{table}
\centering
\caption{Accuracy (percentage of the correctly predicted \rhalf~classes over the total) achieved by different CNN architectures over a limited set of training and validation examples ($4500$ and $500$ light curves respectively). The numbers are for the validation set (training set in parenthesis). The difference between the last two architectures is the inclusion of drop-out (DO) regularization. The best architecture used in the following is highlighted in bold.}
\begin{tabular}{|l||l|l|l|}
Epoch             & 200         & 500         & 1000        \\
\hline
2 blocks          & 30.8 (39.9) & 44.7 (55.4) & 56.9 (61.4) \\
3 blocks          & 58.3 (57.8) & 65.2 (71.0) & 65.2 (77.3) \\
4 blocks          & 51.3 (54.1) & 65.5 (71.2) & 67.4 (79.2) \\
5 blocks		  & 61.7 (96.6) & 70.5 (99.6) & 69.3 (99.8) \\
\textbf{5 blocks (DO)}     & 66.2 (70.6) & 70.3 (81.4) & 75.3 (90.8) \\ \hline
\end{tabular}
\label{tab:acc_model}
\end{table}

The results shown in Table \ref{tab:acc_model} justify the assertion that both deeper models and appropriate regularization have a positive impact on performance.
With respect to the depth of the architecture, more layers-block lead to higher accuracy, noting that 5 blocks perform better for $200$ epochs of training, compared to 2 blocks for $1000$.
Furthermore, we observe that although the performance on the training set reaches $90\%$ when no drop-out regularization is introduced, the corresponding accuracy in the validation set is only $69\%$.
This large gap between training and validation performance is a known problem affecting CNNs called overfitting \citep{Srivastava2014}.
The consequence of this situation is that once the CNN achieves an almost perfect performance on the training set, its performance does not improve any further, thus the generalization capacity of the network is diminished. 
Introducing the drop-out regularization leads to an achieved accuracy for both sets that keeps increasing, as required to avoid overfitting.

\subsection{Validation}
The validation set consists of a part of the training set that was not used during the actual training.
For each \rhalf~class and profile shape, $100$ light curves are set aside as part of the validation set.
The final version of the trained network is then used to predict the class of each of these 50,000 light curves.

\begin{figure*}
	\includegraphics[width=\textwidth]{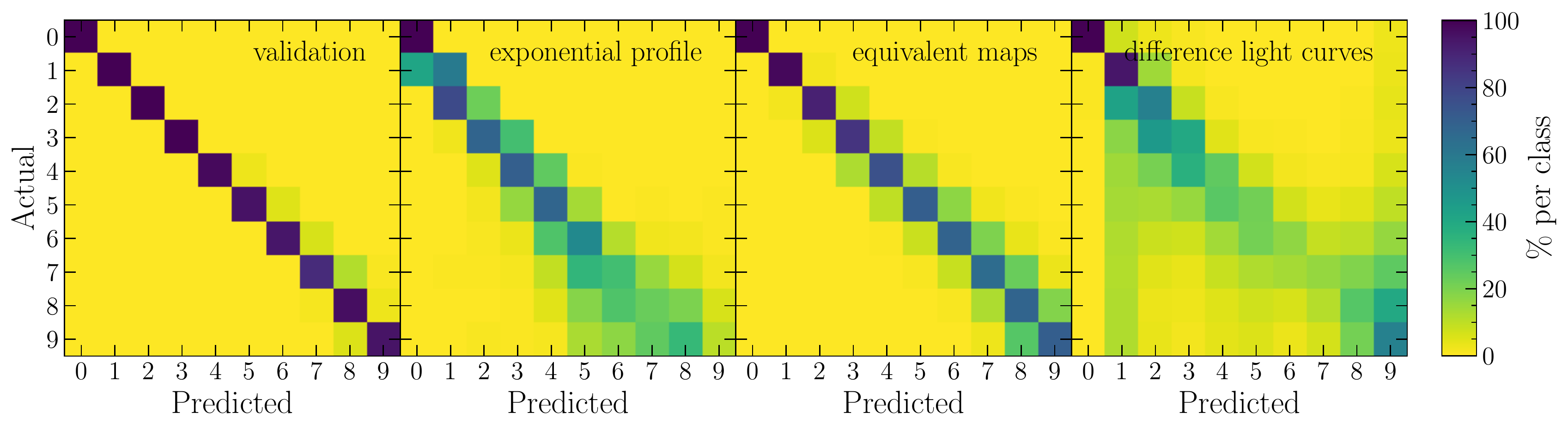}
	\caption{Confusion matrices (actual versus predicted classes of $r_{1/2}$) for the validation set, the `exponential profile', the set of statistically equivalent maps, and for difference light curves (see the corresponding sections in the text for the description of each set). Each row of the matrices is normalized to 100 per cent. The higher the percentage on the diagonal the more accurate the network predictions. The different classes are listed in Table \ref{tab:classes}.}
	\label{fig:confumat}
\end{figure*}

The confusion matrix is a standard way of visualizing the performance of algorithms dealing with classification problems.
For each light curve and regardless of the profile shape, the true value of \rhalf~(the actual class) and the output of the trained network (the predicted class) are stored in a matrix.
If all the entries lie on the diagonal, then all the predicted classes coincide with the actual classes and the network predictions are 100 per cent accurate.
We obtain a 96 per cent accuracy for the validation set, whose confusion matrix is shown in the left panel of Fig. \ref{fig:confumat}.

\subsection{Further tests}
\label{sec:further_tests}
The validation set consists of light curves that are generated from the same underlying magnification map as those used in the training.
Thus, they are expected to have very similar properties, which explains the 96 per cent accuracy of the predictions - the network performs well on the same kind of data that it was trained on.

To push the limits of the predictive power of our network, a number of `unfair' tests was performed using data sets that were designed to be different to the training.
We examined light curves from a profile with a shape unknown to the network - the exponential profile, from different realizations of statistically equivalent maps (having the same $\kappa,\gamma$ as the training but different caustic networks), maps with varying $\kappa,\gamma$ throughout the parameter space, and difference light curves.
In all of these tests the same neural network was used, trained only on light curves generated from the magnification map shown in the top left of Fig. \ref{fig:lcurves_maps} and the profile shapes of Table \ref{tab:shapes}.
We point out once more that the network output is always the class of \rhalf~listed in Table \ref{tab:classes}, regardless of the profile shape.

\subsubsection{Classifying an unknown profile}
\label{sec:expo}
The first test is to classify light curves from a profile whose shape is unknown to the network, i.e. different to the shapes used in the training set.
We used the exponential profile, shown in Table \ref{tab:shapes}.
Because this profile is unbounded for $r \rightarrow 0$, we introduce an artificial cut off at 10 times the brightness value at \rhalf~(at $r\approx 0.2 $ \rhalf) and assume this value for the smaller radii.
Due to this sharp peak at $r \rightarrow 0$, this profile is not expected to smooth out light curve features by much compared to a point source.
Also, this profile is quite different to the ones used in the training which do not have such a sharp peak at low values of $r$.

For each profile size, 1,000 light curves are selected that have a maximum magnification of at least 1.5, leading to a total of 10,000 light curves to be classified.
The confusion matrix, shown in Fig. \ref{fig:confumat}, is mostly diagonal but just underestimates the profile size by 1-3 classes.
The low percentage of light curves classified correctly, which is 30 per cent (on the diagonal), is therefore slightly misleading.

\subsubsection{Statistically equivalent maps}
\label{sec:different_realizations}
We apply our neural network to light curves from 14 different realizations of magnification maps with $\kappa=0.09,\gamma=0.55$, same as the one used to create the training set.
The number of microlenses and the magnification histogram of these maps remain the same, while the random microlens positions are varied to produce different caustic networks.
For each map, profile shape, and size, 100 light curves are selected that have a maximum magnification of at least 1.5, leading to a total of 70,000 mock light curves in this test set (7,000 from each class of \rhalf).
82 per cent of the light curves are classified correctly, lying on the diagonal of the confusion matrix shown in Fig. \ref{fig:confumat}.
Confusion matrices per profile shape for this set of maps are shown in Fig. \ref{fig:app_confumat}.

\subsubsection{Difference light curves}
\label{sec:difference_lc}
In order to obtain microlensing light curves from real observations, one has to subtract the time series of the brightness measurements between pairs of multiple quasar images (after correcting for the time delay).
This removes the (unknown) intrinsic variability of the quasar itself, which is the same in all the multiple images, and the residual signal is attributed to microlensing.
However, this results in a fundamental issue that cannot be addressed: there is no way to know which of the two images is undergoing some microlensing effect and to what extent (or if both are at the same time).
Therefore, to model observational light curves one has to consider such difference light curves.

Difference light curves result by subtracting the brightness in magnitudes (equivalent to dividing the magnification) of two light curves that originated from different quasar images, most likely having quite different $\kappa,\gamma$ parameters.
This means that each difference light curve is now associated with two magnification maps rather than just one.
To create such pairs of maps, we select 5 maps from the set of statistically equivalent maps used previously and combine them to form 10 unique pairs.
It is pointed out that the same accretion disc profile (size and shape) has to be used with both maps before creating the difference light curves.
Due to the combinatorial explosion of the size of the data, 10 of the previously generated light curves per profile shape, size, and map are selected, resulting in 50,000 difference light curves in total (5,000 for each class of \rhalf).

The confusion matrix for this test set is shown in Fig. \ref{fig:confumat}.
The matrix is still mostly diagonal, with 42 per cent of the light curves classified correctly.
We point out that applying this combination and division operation to the data is expected to introduce new features that the network has not been trained to recognize, in a way, creating a new type of data.
Thus, the observed drop in the number of correct predictions is not surprizing.

Finally, for this test we used pairs of maps with the same $\kappa,\gamma$ values, although real pairs of quasar images are expected to have quite different ones.
As it turns out, the effect of varying $\kappa,\gamma$ is much more prominent even without considering difference light curves, and is examined separately below.

\subsubsection{Parameter space}
\begin{figure}
	\includegraphics[width=0.5\textwidth]{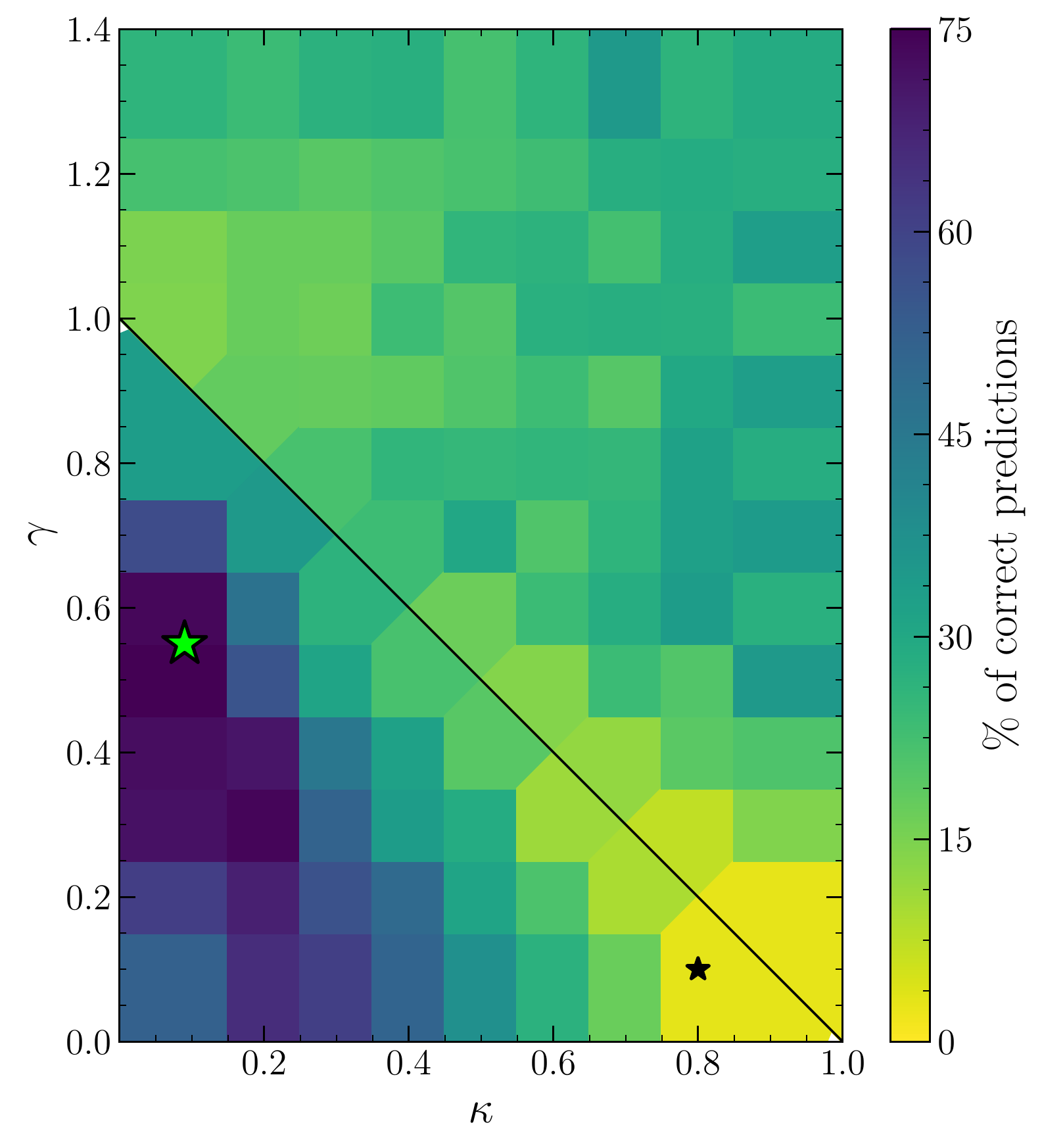}
	\caption{Percentage of correct size predictions from 108 magnification maps throughout the $\kappa,\gamma$ parameter space (each map location is represented by a Voronoi cell). The closer to the $\kappa,\gamma$ location used in the training (green star, top map in Fig. \ref{fig:lcurves_maps}), the higher the percentage of correct predictions. Light curves become more different the closer the maps get to the critical line (black diagonal line), where the macro-magnification $\mu_{\rm macro} = [(1-\kappa)^2 - \gamma^2]^{-1}$ diverges. The location of the bottom map in Fig. \ref{fig:lcurves_maps} is also indicated (black star).}
	\label{fig:pspace}
\end{figure}

For this test, a set of 108 magnification maps is selected throughout the $\kappa,\gamma$ parameter space, shown in Fig. \ref{fig:pspace}.
For each map, profile shape, and size, 200 light curves are selected that have a maximum magnification of at least 1.5, resulting in a total of $\approx10^6$.
The confusion matrices are calculated in each case, but are hard to visualize in the $\kappa,\gamma$ parameter space.
Instead, the percentage of the correct predictions is computed for each map (the confusion matrix trace), for all the profile shapes, and shown in Fig. \ref{fig:pspace}.

The highest percentage of correct predictions is between 70 to 80 per cent, and it occurs near $\kappa=0.09,\gamma=0.55$, the values that the network was trained on.
The network performs well in a wide area around these values, which is not surprizing because the caustic networks and the statistical properties of the maps are quite similar there \citep[see fig. 4 of][for the magnification histograms in the parameter space]{Vernardos2013}.

The network performs the worst near the critical line (black line on Fig. \ref{fig:pspace}) for high values of $\kappa$.
The critical line is where two macro-images of the quasar merge or annihilate and the macro-magnification $\mu_{\rm macro} = [(1-\kappa)^2 - \gamma^2]^{-1}$ diverges.
The number of microlenses dramatically increases and caustic networks become denser and superimposed, leading to quite different light curves.
The magnification map shown in the bottom panel of Fig. \ref{fig:lcurves_maps} lies much closer to the critical line and the difference with the map used in the training is striking: there are several and constant small variations in the light curve as the source is continuously traversing dense caustic networks, as opposed to simple caustic crossing events displaying the characteristic double peak as the source enters and exits an isolated diamond caustic.

\subsection{Recovering accretion disc parameters}
Measuring the accretion disc size in a range of wavelengths can constrain the free parameters in equation (\ref{eq:parametric}), i.e. the size $r_0$ and slope $\nu$.
Fixing $r_0$ and $\nu$ to some fiducial values, in this case $2.59\times10^{15}$ cm and $4/3$, matches a wavelength to each accretion disc size (\rhalf).
We can then use the sizes predicted in each wavelength by the neural network to infer the underlying $r_0$ and $\nu$ values.

However, we are dealing with what is effectively a pattern recognition problem and any kind of physical scale is irrelevant; stretching or compressing the light curve data in the length or magnification directions does not affect the network predictions, as long as the \rhalf/\rein~and pixels/\rein~ratios, which determine the relevant size of the profiles and the light curve resolution, remain fixed.
Thus, the actual values of $r_0$ and $\nu$ used to match sizes to wavelengths are actually irrelevant.

For each trajectory on the magnification map we extract the light curve data of a given accretion disc profile shape for all values of \rhalf.
Predicting the size of each light curve in such collections leads to a measurement of \rhalf~as a function of wavelength.
Converting these measurements to logarithmic space and performing a simple linear fit leads to a measurement of $r_0$ and $\nu$.

\begin{figure}
	\includegraphics[width=0.5\textwidth]{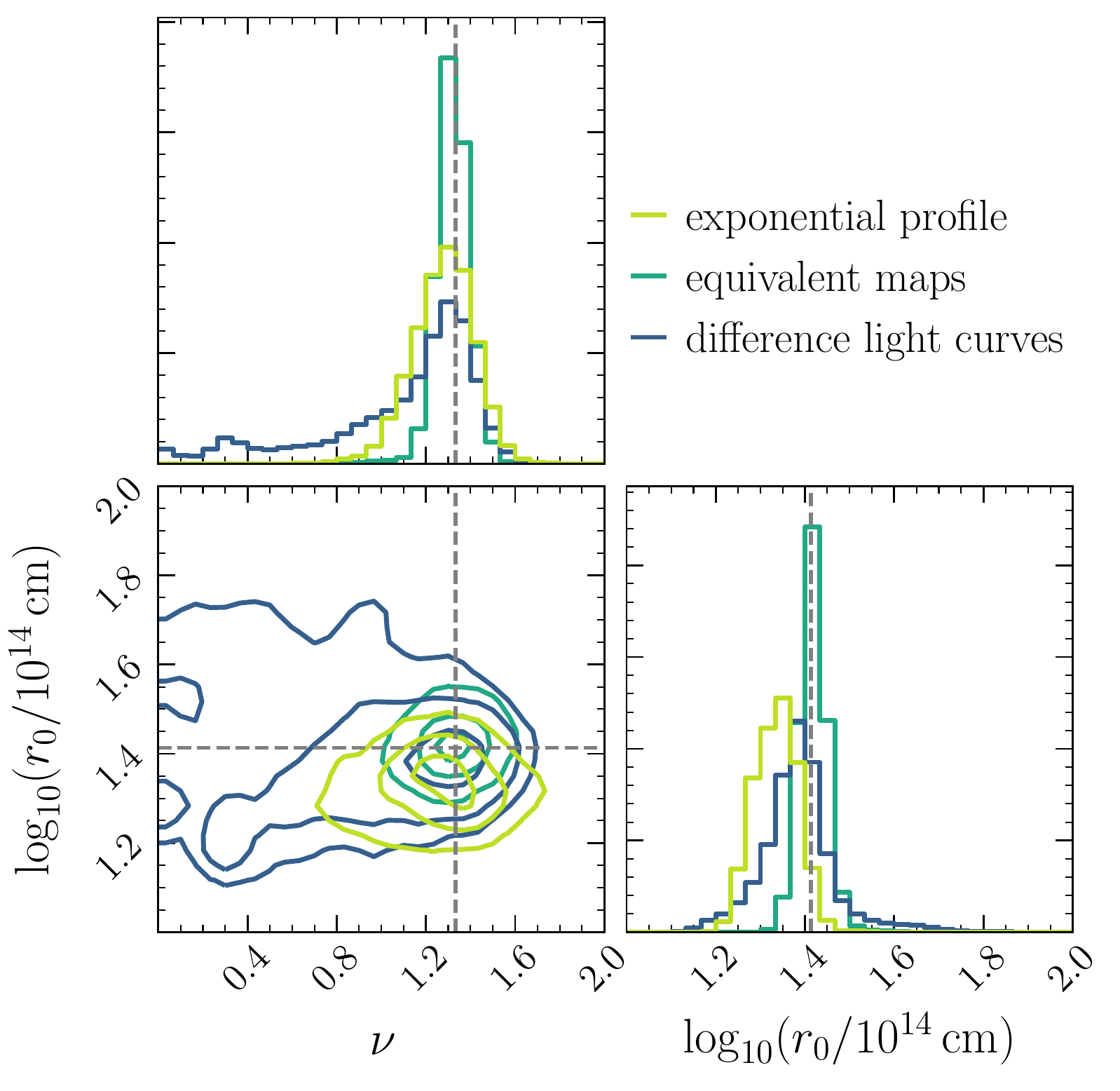}
	\caption{Probability density and histograms for the accretion disc parameters $r_0$ and $\nu$ (see equation \ref{eq:parametric}). Measurements are obtained from the exponential profile, equivalent maps, and difference light curve datasets. The shape of the probability density surface depends purely on the ability of the network to correctly predict the accretion disc size, and is related to the confusion matrices of these datasets, shown in Fig. \ref{fig:confumat}. Dashed lines indicate the true underlying values used in this case, i.e. $\nu=4/3$ and $r_0 = 2.59 \times 10^{15}$ cm. Contours are drawn at the 68, 95, and 99 per cent confidence intervals.}
	\label{fig:slopes}
\end{figure}

We use the test sets for an exponential profile, statistically equivalent maps, and difference light curves created in Section \ref{sec:further_tests}, which provide 1,000, 7,000 and 5,000 measurements of \rhalf~vs wavelength.
The resulting values for $\nu$ and $r_0$ are shown in Table \ref{tab:disc_measurements}.
The probability density contours and histograms are shown in Fig. \ref{fig:slopes}.

\begin{table}
	\centering
	\caption{68 per cent confidence intervals for the accretion disc profile parameters $\nu$ and $r_0$, corresponding to the probability histograms shown in Fig. \ref{fig:slopes}.}
	\begin{tabular}{|r|l|l|}
		dataset           		& $\nu$       					& $r_0$ [$10^{15}$ cm]		\\
		\hline
		exponential	profile		& $1.28_{-0.13}^{+0.11}$		& $2.19_{-0.24}^{+0.20}$	\\
		equivalent maps			& $1.33_{-0.04}^{+0.04}$		& $2.57_{-0.06}^{+0.06}$ 	\\
		difference light curves	& $1.22_{-0.56}^{+0.15}$		& $2.4_{-0.31}^{+0.23}$ 	\\
		\hline
	\end{tabular}
	\label{tab:disc_measurements}
\end{table}

It is pointed out that the shapes of the probability contours in Fig. \ref{fig:slopes} and the resulting fractional errors on the derived $r_0$ and $\nu$ parameters do not depend on their absolute values and only reflect the extent to which the network fails to predict the correct sizes in each case (reflecting the confusion matrix of each test set).
The measurements of $r_0$ and $\nu$ from the validation set (not shown) are therefore expected to be even closer to the true underlying values.

\section{Conclusions}
\label{sec:conclusions}
We used a machine learning approach based on a Convolutional Neural Network (CNN) to study simulated quasar microlensing light curves for the first time.
Our algorithm was trained on and applied to hundreds of thousands of such light curves.
We demonstrated that our method is able to recover the correct size of the underlying accretion disc at 96 per cent accuracy for the validation set, and at 82 for a test set of statistically equivalent light curves, generated from magnification maps with the same $\kappa,\gamma$ values but different random microlens positions.
When light curves from a profile whose shape is unknown to the network are used (the `exponential' profile), the accuracy drops at 15 per cent but the corresponding confusion matrix shown in Fig. \ref{fig:confumat} remains almost diagonal (the network underpredicts the sizes by 1-3 classes).
These two tests ensure that the training data are not overfitted.

A further test was made with difference light curves - combinations of light curves from two different magnification maps.
The accuracy dropped to 42 per cent, which is an acceptable performance given that difference light curves introduce new features to the data that the network has not been trained on.

We find that the specific shape of the accretion disc profile has a negligible effect on microlensing properties of light curves, supporting and extending the results of \cite{Mortonson2005}.
This is demonstrated by the almost diagonal confusion matrices shown in Figs. \ref{fig:confumat} and \ref{fig:app_confumat}.
However, on small scales (short length/time), where isolated caustic crossing events can be observed, the detailed structure of the accretion disc still makes a difference (see inset plot in Fig. \ref{fig:lcurves_maps}).

Classifying accretion disc sizes in many wavelengths leads to measurements of the accretion disc temperature profile via equation (\ref{eq:parametric}).
Fig. \ref{fig:slopes} shows the probability of such measurements for thousands of light curves from three different test sets.
The errors on recovering the correct underlying structure parameters depend purely on the accuracy of the network size predictions.
In particular, for the `exponential' profile shape the absolute size predictions are well-off the truth, as expected from the confusion matrix, but the relative scaling, and hence the power-law index $\nu$, is correctly recovered.

Our network was trained on a single magnification map with $\kappa=0.09,\gamma=0.55$ and then applied to the entire $\kappa,\gamma$ parameter space shown in Fig. \ref{fig:pspace}.
Predictions near the training map location are systematically higher than elsewhere in the parameter space, indicating that there is a potentially measurable effect of the $\kappa,\gamma$ values on the light curves.
This could lead to measurements of the smooth matter fraction - constraining the partition between smooth (dark) and compact (stellar) matter - which are crucial to lifting the degeneracy between dark matter and the initial mass function in the lens \citep[e.g.][]{Oguri2014}.

The focus of this work is to study light curves in an abstract way, taking into account only the size of the accretion disc with respect to \rein.
Thus, light curve data were assumed to be long (5$\times$\rein), continuous, and containing some minimum amount of information (all light curves were selected to have a maximum magnification of 1.5, guaranteeing at least some variations).
The most important caveat is the length of the light curves, which determines the impact of other factors not considered here, like the effective velocity model.
However, the applicability of this new technique to real data will ultimately depend on how well observational effects, like sparse data, gaps, irregular sampling, etc, can be taken into account and mitigated.

In this work we demonstrated the feasibility of analyzing hundreds of thousands of microlensing light curves.
This method opens up a whole new range of possibilities of quantifying light curve properties systematically and consistently throughout the microlensing parameter space.
This is important for the upcoming all-sky surveys like LSST, which will provide data on thousands of microlensed quasars.

This new CNN approach was explored along two perpendicular directions: improving the network predictions on a restricted training set by modifying its architecture, and then applying it to a wider range of data.
The next step is to couple them, include observational effects, and improve the predictions on all the new architecture/data possibilities (difference light curves, throughout the $\kappa,\gamma$ parameter space, as a function of light curve length, etc).
The new technique presented here brings us closer to addressing the pertinent question: ``At which point all the available information has been extracted from a quasar microlensing light curve?''.
Eventually, the most critical parameter will be the light curve length, which is necessary for designing future observations, follow-up campaigns, etc.

\section*{Acknowledgements}
GV is supported through an NWO-VICI grant (project number 639.043.308).
The authors would like to thank the referee James H. H. Chan for his comments that improved the final version of this work.


\bibliographystyle{mnras}
\bibliography{main} 


\appendix

\section{Confusion matrices per profile shape}
\begin{figure*}
	\includegraphics[width=\textwidth]{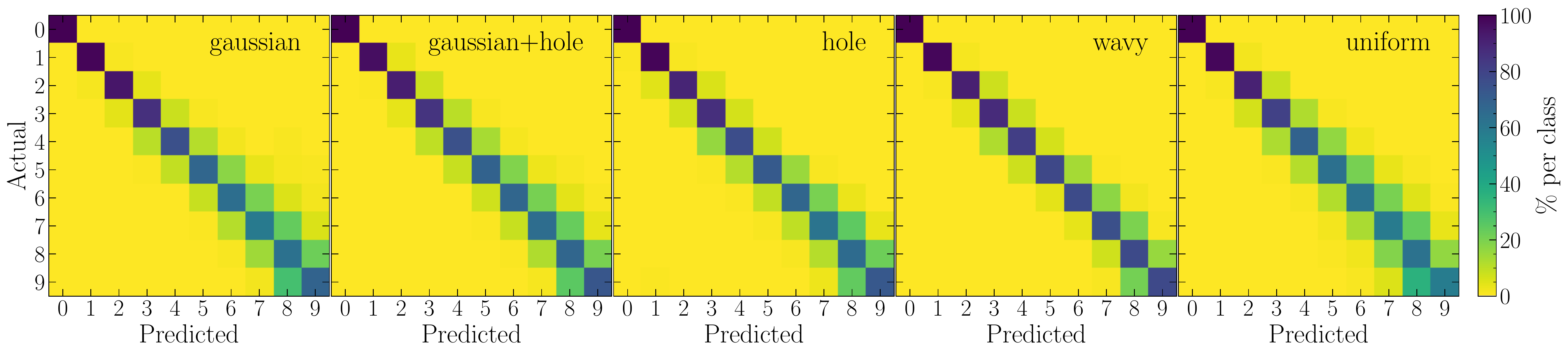}
	\caption{Confusion matrices (actual versus predicted classes of $r_{1/2}$) per profile shape for the set of statistically equivalent maps. The middle panel of Fig. \ref{fig:confumat} is the sum of these constituent confusion matrices. Each row of the matrices is normalized to 100 per cent. The higher the percentage on the diagonal the more accurate the network predictions. The different classes are listed in Table \ref{tab:classes}.}
	\label{fig:app_confumat}
\end{figure*}


\bsp	
\label{lastpage}
\end{document}